\documentclass[12pt]{article}
\usepackage{chicago}

\usepackage{natbib}
\usepackage{lscape}
\usepackage{geometry}
\usepackage{booktabs}
\usepackage{graphicx}

\begin{document}

\title{Dynamic Model Averaging in Large Model Spaces Using
Dynamic Occam's Window\thanks{%
The views expressed are those of the authors and do not necessarily
reflect those of the Central Bank of Ireland.}}
\author{Luca Onorante\thanks{%
Luca Onorante is Head of the Macro Modelling Project and Deputy Head of Research, Central Bank of Ireland.} \and Adrian E. Raftery\thanks{Adrian E. Raftery is Professor of Statistics and Sociology at the University of Washington.}}
\date{\today}
\maketitle

\begin{abstract}
Bayesian model averaging has become a widely used approach to accounting for 
uncertainty about the structural form of the model generating the data.
When data arrive sequentially and the generating model can change over time,
Dynamic Model Averaging (DMA) extends model averaging 
to deal with this situation. Often in macroeconomics, however, 
many candidate explanatory variables are available and the number of 
possible models becomes too large for DMA to be applied in its original form.
We propose a new method for this situation which allows us to perform DMA
without considering the whole model space, but using a subset of models
and dynamically optimizing the choice of models at each point in time.
This yields a dynamic form of Occam's window.
We evaluate the method in the context of the problem of nowcasting
GDP in the Euro area. We find that its forecasting performance compares
well that of other methods. 

\bigskip

\end{abstract}

\newpage
\baselineskip=18pt

\section{Introduction}
In macroeconomic forecasting, there are often many available candidate
predictor variables, and uncertainty about precisely which ones should
be included in the forecasting model. Accounting for this uncertainty
is important, and Bayesian model averaging (BMA)
\citep{Leamer1978,Raftery1988,MadiganRaftery1994,Raftery1995,Raftery&1997,Hoeting&1999}
has emerged as an established way of doing so, thanks in large part
to methodological developments to which Eduardo Ley has contributed
\citep{FernandezSteel2001,Fernandez&2001,LeySteel2007a,LeySteel2007b}.
For other developments of BMA in macroeconomics, 
see \citet{BrockDurlauf2001}, \citet{Brock&2003}, \citet{Sala&2004}, 
Durlauf et al (2006, 2008), \nocite{Durlauf&2006,Durlauf&2008} 
\citet{Eicher&2010} and \citet{Varian2014}; see \citet{Steel2011} for a survey.

Model averaging has been found useful in economics for several reasons.
These include the possibility of using more parsimonious models,
which tend to yield more stable estimates, because fewer degrees of freedom are 
used in individual models. Also, BMA yields the posterior inclusion
probability for each variable, making the results easy to interpret.
Further, when compared to large scale models with the same variables,
BMA provides automatic shrinkage through the parsimony of the individual
models. It can be used to account for uncertainty about the best way 
to measure a concept, such as different measures of slack in a 
Phillips curve.
It can also be used to account for uncertainty about
model structure beyond variable selection (e.g. linear versus nonlinear,
univariate versus multivariate, fixed versus time-varying parameters).

BMA is a static method, and as such cannot deal with the situation where
data arise sequentially (as is generally the case in macroeconomics),
and the form of the generating model can change over time.
To deal with this, Dynamic Model Averaging (DMA) was proposed
by \citet{Raftery&2010}. This allows both the model form and the model
parameters to change over time and tracks both recursively.

DMA requires the evaluation of each model at each time point.
Often in macroeconomics, however,
many candidate predictor variables are available.
Typically, the number of candidate regression models is equal or
close to $2^J$, where $J$ is the number of candidate predictor variables.
This grows rapidly with $J$, and so the number of
possible models becomes too large for it to be computationally
feasible for DMA to be applied in its original form.

We propose a way of implementing DMA in large model spaces, called
Dynamic Occam's Window (DOW). This allows us to perform DMA
without considering the whole model space, but using a subset of models
and dynamically optimizing the choice of models at each point in time.
It is particularly adapted to macroeconomic studies and allowing the inclusion of big information sets. Our proposal allows us to run DMA without an exhaustive exploration of the space of models.
We describe an application to the difficult problem of nowcasting GDP in the euro area.

The paper is organized as follows. 
Section 2 briefly reviews Bayesian and Dynamic Model Averaging and 
Model Selection. In Section 3 we describe the Dynamic Occam's Window method.
In Section 4 we describe an economic application, the nowcasting of the 
euro area GDP, and show in Section 5 that the results of our technique 
compare well with others in the existing literature in terms of 
forecasting performance and computational efficiency. 
Section 6 gives the results of several sensitivity analyses, 
and the final section concludes.

\section{Forecasting with Dynamic Model Averaging}

Here we outline the main concept of Dynamic Model Averaging (DMA),
introduced by \citet{Raftery&2010}.
Assume a population, $M$, of $K$ candidate regression models, 
$M = \{m_1,..,m_K\}$, where model $m_k$ takes the form:

\begin{equation}
y_{t} = x_{t}^{\left( k\right) }\beta _{t}^{\left( k\right) }+\varepsilon
_{t}^{\left( k\right) }  \label{model_space} \\
 ,\qquad  \nonumber
\end{equation}%
where $\varepsilon _{t}^{\left( k\right) } \sim N\left( 0,\sigma
_{t}^{2\left( k\right) }\right) $ . 

Each explanatory set $x_{t}^{\left( k\right) }$ contains a subset of the potential
explanatory variables $x_{t}$. It can be seen immediately that this implies a large number 
of models: if $J$ is the number of explanatory
variables in $x_{t}$ there are $K=$ $2^{J}$ possible regression models 
involving every possible combination of the $J$ explanatory variables. 

DMA averages across models using a recursive updating scheme. 
At each time two sets of weights are calculated, $w_{t|t-1,k}$ and $w_{t|t,k}$.
The first, $w_{t|t-1,k}$, is the key quantity. 
It is the weight of model $k$ in 
forecasting $y_{t}$ given data available at time $t-1$. 
The second weight, $w_{t|t,k}$, is the update of $w_{t|t-1,k}$ 
using data available at time $t$. DMA produces forecasts
which average over all $K$ models using $w_{t|t-1,k}$ as weights.
Note that DMA is dynamic since these weights can vary over time. 

Dynamic Model Selection (DMS)  is similar but it involves selecting the model
with the highest value for $w_{t|t-1,k}$ and using it for forecasting $y_{t}$%
. 
DMS allows for model switching: at each point in time it is
possible that a different model is chosen for forecasting. 

\citet{Raftery&2010} derive for DMA the following updating equation:
\begin{equation}
w_{t|t,k}=\frac{w_{t|t-1,k}L_{k}\left( y_{t}|y_{1:t-1}\right) }{%
\sum_{\ell=1}^{K}w_{t|t-1,\ell}L_{\ell}\left( y_{t}|y_{1:t-1}\right) } ,
\label{modupdate}
\end{equation}%
where $L_{k}\left( y_{t}|y_{1:t-1}\right) $ is the predictive
likelihood, or the predictive density of $y_{t}$ for model $m_k$ evaluated 
at the realized value of $y_{t}$. The
algorithm then produces the weights to be used in the following period by 
using a forgetting factor, $\alpha $:
\begin{equation}
w_{t+1|t,k}=\frac{w_{t|t,k}^{\alpha }}{\sum_{\ell=1}^{K}w_{t|t,\ell}^{%
\alpha }}.  \label{modpred}
\end{equation}
The forgetting factor, $\alpha$, is specified by the user.
Here we use $\alpha=0.99$, following Raftery et al. (2010).

Thus, starting with $w_{0|0,k}$ (for which we use the noninformative choice
of $w_{0|0,k}=\frac{1}{K}$ for $k=1,..,K$) we can recursively calculate the
key elements of DMA: $w_{t|t,k}$ and $w_{t|t-1,k}$ for $k=1,..,K$.

\section{Dynamic Occam's Window}

When many potential regressors are considered, the number of models is too high to be tractable. However, typically the great majority of models contribute little to the forecast, as their weights are close to zero. These include for example highly misspecified models, which are kept despite their poor performance only to calculate equation (2).

We propose a heuristic aiming at eliminating most of these low probability models from the computation, while being able to ``resurrect'' them when needed. 
This is based on the Occam's Window method of \citet{MadiganRaftery1994}, 
in which model averaging is based only on models whose posterior 
model probability is greater than some multiple $C$ of the highest
model probability. (\citet{MadiganRaftery1994} used $C=1/20$, while 
subsequent implementations have used lower values.) 

We now extend Occam's Window to the dynamic context.
Our Dynamic Occam's Window (DOW)  method is based on two implicit assumptions:

\begin{enumerate}
\item We dispose at the initial time of a valid population of models 
\item Models do not change too fast over time: the relevant models at each time are close enough (in a ``neighbourhood'' opportunely defined) to those of the preceding time.
\end{enumerate}

We believe these assumptions are reasonable in typical problems of macroeconomic analysis. If verified, they allow the exploration of the space of models in a parsimonious and efficient way.

\subsection{Forecast, Expand, Assess, Reduce: the FEAR algorithm}

We propose to implement Occam's window on currently used models and 
keep for future use only those that perform sufficiently well relative 
to the best performer. Call the current set of models $M_0(t)$,
and renormalize their current weights, $w_{t|t,k}$, so that 
they sum to 1 over the current set of models, i.e. so that
$\sum_{k:m_k \in M_0(t)} w_{t|t,k} = 1$.
After choosing a threshold $C \in (0,1]$, we keep for future use the models 
$m_k \in M_0(t)$ that fall in Occam's window, namely
\begin{equation}
m_k: m_k \in M_0(t), w_{t|t,k} \geq 
  C * \max_{\ell:m_\ell \in M_0(t)} w_{t|t,\ell}.
\label{eq-OW}
\end{equation}

The FEAR algorithm iterates four steps: Forecasting, Expanding the set of models, Assessing them, and Reducing the model set via Occam's window. 
\bigskip

Initialization

\begin{enumerate}
\item Divide the sample $1, \ldots , T$ into a training sample 
$1, \ldots , T_r$ and a forecasting/evaluation sample $T_{r+1}, \ldots , T$

\item Start with an initial population of models $M_0(T_r)$ and an 
initial set of weights $w_{T_r|T_r,k}$.
\end{enumerate}

For $t=(T_r+1), \ldots , T$:

\begin{enumerate}

\item (Forecast) Use the models in $M_0(t-1)$ and the weights $w_{t|t-1,k}$ 
to perform model averaging as in \citet{Raftery&2010}, 
obtaining the forecast distribution $p(y_t | y_{1:t-1}, M_0(t-1))$.

\item (Expand) Expand $M_0(t-1)$ into a larger population $M_1(t)$ including all $m_k \in M_0(t-1)$ and all their neighbouring models (all models derived from any model $m\in M_0(t-1)$ by adding or subtracting a regressor).

\item (Assess) Upon observing $y_t$ compute for all $m_k \in M_1(t)$ the
weights $w_{t|t,k}$, normalized to sum to 1 over $M_1(t)$.

\item (Reduce) Define the final population of models for time $t$, $M_0(t)$,
as those in $M_1(t)$ that are in Occam's Window, namely
$M_0(t) = \{m_k \in M_1(t): w_{t|t,k} \geq C *  \max_{\ell:m_\ell \in M_1(t)} w_{t|t,\ell} \}$.
\end{enumerate}
End for

\subsection{Computational issues}

This section explains why Dynamic Occam's Window allows the exploration of large
models spaces that would not be possible otherwise.

We define, rather imprecisely but as a rough reference, a Notional Unit of Computation (NUC) as a basic operation of estimation. Since we are concentrating on computability, we consider broadly equivalent (one NUC) one OLS estimation, one period estimation of a Kalman filter and in general each operation involving at least a matrix inversion. On this loosely defined but quite general metric we compare the DOW method with a DMA exhaustively exploring  the space of models.
Let $J$ be the number of candidate explanatory variables,
$T$ be the number of time points for which we have data, and
$N$ be the number of models in Occam's window (a subset of the $K$ possible regression models).

DMA with all models has approximate computational cost
\begin{equation}
{\rm NUC}_{\rm DMA} = 2 ^{J} * T ,
\label{eq-nucdma}
\end{equation}
because all the potential models need to be estimated once per period.

The DOW method reduces the number of models to be evaluated but changes the population dynamically. It is therefore necessary to re-estimate each model from the beginning each time. Its computational cost is thus approximately
that of estimating about
\begin{equation}
{\rm NUC}_{\rm DOW} = \frac{(T + 1 )* T }{2} * N
\label{eq-nucdow}
\end{equation}
different models. The role of the number of models $N$ is explored in Section 6. 

The DOW method allows gains in speed when ${\rm NUC}_{\rm DOW} < {\rm NUC}_{\rm DMA}$, or
\begin{equation}
N < 2 * \frac{2^{J}}{(nT + 1)} .
\end{equation}

To illustrate, consider the case where $T$ = 45, $J$ = 25, and
$N$ varies.  Then for $N=1000$
\begin{equation}
\frac{{\rm NUC}_{\rm DOW}}{{\rm NUC}_{\rm DMA}} = \frac{10.350.000}{1.509.949.440} = 0.68\% .
\end{equation} 
Thus the DOW method is about 150 times faster. 

\begin{figure}
   \centering
       \includegraphics[width=1\textwidth]{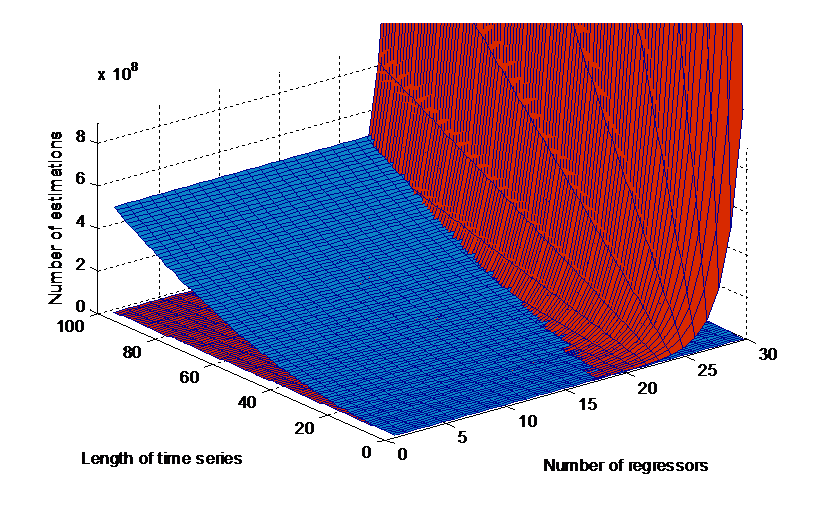}
   \caption[Computing Time] 
  {\label{fig-computingtime} Computing Time:
The number of NUC (vertical axis) plotted against the data length and the number of regressors. The blue area refers to the DOW method, the red area to DMA.}
 \end{figure}
 
Figure \ref{fig-computingtime} shows the relationships (\ref{eq-nucdma}) and 
(\ref{eq-nucdow}).
The computational complexity of the DOW method grows quadratically with the length of the available series $T$, while that of DMA grows only linearly in $T$ but increases exponentially in the number of regressors $J$. Above 15-20 regressors the DOW method is always more efficient computationally. This is particularly true when the time series are relatively short, since longer series imply a higher number of estimations for each model in the case of Occam's window.

\section{An Economic Application: Forecasting GDP in the Euro Area
in the Great Recession}

\subsection{The Forecasting Application}
Short term forecasting and nowcasting economic conditions is important for policy makers, investors and economic agents in general. Given the lags in compiling and releasing key macroeconomic variables, it is not surprising that particular attention is paid to nowcasting, an activity of importance because it allows economic decisions to be made and policy actions to be taken with a more precise idea of the current situation. 

We apply the DOW method to nowcasting GDP in the euro area. This problem is particularly difficult because there are many candidate explanatory variables (large $J$) but most of them cover a short time span (small $T$). We use quarterly (or converted to quarterly) series available from 1997, and we describe our source data in Table \ref{tbl-variables}.\footnote{We use quarterly data or data at higher frequency. Higher frequency data are converted into quarterly by taking the last observation (for example the last month in a quarter). As a robustness check we also experimented with averages.} Abstracting from minor differences in publication dates, there are two main GDP nowcasts that a forecaster may perform, depending on whether or not the preceding quarter figure for GDP is available. For simplicity of exposition, we focus on the case when the past quarter is already available. Our nowcasts will be based on an information set comprising past GDP and current indicators. 

\begin{landscape}
\begin{table}
\begin{center}
\begin{tiny}
\caption{\label{tbl-variables} Variables Used in the Euro Area GDP 
Forecasting Area Application}
\begin{tabular}{@{}llll@{}}
\toprule
CODE         & VARIABLE
& TRANSFORMATION     & DATA SOURCE \\ \midrule
hicp         & Euro area - HICP, Eurostat
& Annual growth rate & Eurostat    \\
hicpx        & Euro area - HICP excl. unprocessed food an energy, Eurostat
& Annual growth rate & Eurostat    \\
ppi          & Euro area - Producer Price Index, domestic sales, Total Industry (excluding construction) - Neither
seasonally nor working day adjusted      & Annual growth rate & Eurostat    \\
unemp        & Euro area - Standardised unemployment, Rate, Total (all ages), Total (male \& female), percentage of
civilian workforce                      & Level              & Eurostat    \\
ip           & Euro area - Industrial Production Index, Total Industry (excluding construction) - Working day
adjusted, not seasonally adjusted             & Annual growth rate & Eurostat    \\
l1 to l5     & Euro area - Gross domestic product at market price, Chain linked, ECU/euro, Working day and
seasonally adjusted                              & Annual growth rate & Eurostat    \\
spfgdp2      & Survey Professional Forecasters; Forecast topic: Real GDP; Point forecast
& Annual growth rate & ECB         \\
spfgdp2\_var & Survey Professional Forecasters; Forecast topic: Real GDP; Variance of Point forecast
& Variance           & ECB         \\
pmi\_e       & PMI employment
& Level              & Markit      \\
pmi\_ord     & PMI new orders
& Level              & Markit      \\
pmi\_y       & PMI output
& Level              & Markit      \\
x\_oil       & Brent crude oil 1-month Forward - fob (free on board) per barrel - Historical close, average of
observations through period - Euro           & Annual growth rate & ECB         \\
x\_rawxene   & Euro area , ECB Commodity Price index Euro denominated, import weighted, Total non-energy commodity
& Annual growth rate & ECB         \\
x\_neer      & Nominal effective exch. Rate
& Annual growth rate & ECB         \\
xusd         & Exchange Rates; Currency: US dollar; Currency denominator: Euro; Exchange rate type: Spot
& Annual growth rate & ECB         \\
i\_short     & Euro area - Money Market - Euribor 1-year - Historical close, average of observations through period
- Euro                                  & Level              & Reuters     \\
i\_long      & Euro area - Benchmark bond - Euro area 10-year Government Benchmark bond yield - Yield - Euro
& Level              & ECB         \\
m3           & Euro Area - Monetary aggregate M3, All currencies combined - Working day and seasonally adjusted
& Annual growth rate & ECB         \\
spread       & Government bond, nominal yield, all issuers whose rating is triple A (less) Government bond, yield
nominal, all issuers all ratings included & Level              & ECB         \\
stress       & Composite Indicator of Systemic Stress;  Euro area ; Systemic Stress Composite Indicator
& Level              & ECB         \\
risk\_eb     & Euro area , Financial market liquidity indicator: Foreign currency, equity and bond markets
& Level              & ECB         \\
risk\_tot    & Euro area , Financial market liquidity indicator: Composite indicator
& Level              & ECB         \\
risk\_glob   & Euro area , Financial market liquidity indicator: Global risk aversion indicator
& Level              & ECB         \\
risk\_mon    & Euro area , Financial market liquidity indicator: Money market
& Level              & ECB         \\
stox         & Dow Jones Eurostoxx 50 Index - Historical close, average of observations through period
& Annual growth rate & Reuters     \\
domcred      & Euro area , Loans {[}A20{]} and securities {[}AT1{]}, Total maturity, All currencies combined -
Working day and seasonally adjusted          & Annual growth rate & ECB         \\ \bottomrule
\end{tabular}
\end{tiny}
\end{center}
\end{table}
\end{landscape}

The need to use timely indicators largely dictates the choice of potential regressors, but most sectors and economic concepts are well covered. Our indicators include domestic prices (HICP, HICP excluding food an energy and producer prices), cycle indicators (unemployment rate, industrial production, lags of GDP), expectations (mean and dispersion of 2 years-ahead SFP forecasts for GDP, PMI for employment, orders and output), prices of commodities (oil prices, non-energy commodity prices), exchange rates (nominal effective exchange rate, EUR/USD exchange rate), monetary policy variables (short and long interest rates, M3), financial variables (spread between interest rate on bonds of AAA states and average interest rate on bonds, Dow Jones Eurostoxx index, domestic credit). Given the relevance of uncertainty in the macroeconomic developments included in our sample, we also include potential macroeconomic risk indicators (Composite Indicator of Systemic Stress, Risk Dashboard data on banking, total, global and monetary factors). 

All variables are on the year-on-year growth rate scale, with the exception of interest rates and indicators. The target variable in our forecasting exercises is the year-on-year GDP growth rate. As a result, at least four lags of the independent variable must be included as potential regressors; we use five to account for potential autocorrelation in the residuals. This may be overcautious, but unnecessary lags will be selected away in the model averaging. The possibility of adding regressors but discarding them adaptively if they are unnecessary is one of the advantages of our methodology.

In order to concentrate on the effects of the proposed DOW method, we simplify the method  of \citet{Raftery&2010} slightly, and estimate each model recursively but with fixed parameters. We choose this setup because \citep{KoopKorobilis2012} have shown that DMA is a good substitute for time varying parameters, and we want to concentrate on the advantages of the DOW method alone in accounting for model changes. DMA is performed as in \citet{Raftery&2010}, using a discount factor set at $\alpha = 0.99$.

\subsection{Forecasting Performance}
\begin{figure}
   \centering
       \includegraphics[width=1\textwidth]{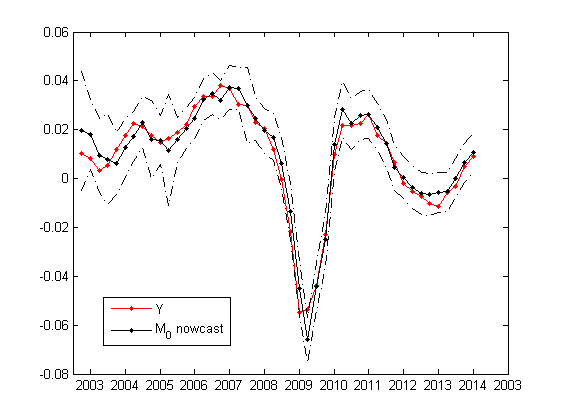}
   \caption{\label{fig-EuroGDP} Euro Area GDP: Nowcasting and uncertainty bands using the DOW method}
 \end{figure}

Figure \ref{fig-EuroGDP} shows that the DOW method had a satisfactory nowcasting performance overall, even in the presence of turning points. The accuracy of the method, as expected, increased with the available data. The 95 per cent error bands take into account the within and between model uncertainty. 

The difficult episode of the recession in 2008-2009 is well captured by DMA. The forecast slightly underpredicts in the trough, but it immediately recovers and becomes quite accurate in the aftermath of the crisis.

Table \ref{tbl-performance} compares the forecasting performance in a pseudo-real time exercise.\footnote{RMSE, MAE and MAX are calculated from numbers expressed in percentages, where 0.01 is one percent.} Practically all the indicators we use are seldom or never revised, the main difference with a real time forecasting exercise being the fact that we use the latest available vintage for GDP. The evaluation sample ranges from 2003q1 until 2014q1.

\begin{table}[h]
   \caption{\label{tbl-performance} 
Forecasting Performance of Different Forecasting Methods}
\begin{center}
\begin{tabular}{@{}lllllll@{}}
\toprule
     & RW   & AR2  & DMA-R& DMA-E& DMS-R& DMA-E    \\ \midrule
RMSE & 0.0101 & 0.0088 & {\bf 0.0043} & {\bf 0.0043} &  0.0048 & 0.0048 \\
MAE  & 0.0067 & 0.0059 & {\bf 0.0033} & {\bf 0.0033} &  0.0035 & 0.0035 \\
MAX  & 0.0332 & 0.0376 & {\bf 0.0125} & 0.0126 &  0.0139 & 0.0139 \\ \bottomrule
\end{tabular}
\end{center}
Note: Methods: RW = Random walk model; AR2 = second-order autoregressive model;
DMA-R = reduced DMA method; DMA-E = expanded DMA method;
DMS-R = reduced DMS method; DMS-E = expanded DMS method. \\
Metrics: RMSE = root mean squared error; MAE = mean absolute error;
MAX = maximum absolute error. The best method by each metric is shown in
bold font. 
\end{table}

The DMA-R forecast is based on the smaller population of models $M_0$ and compares favourably with simple benchmarks. It largely beats both the simple random walk and a standard $AR(2)$. We recall that the forecast DMA-R is based on past GDP and recent information on the indicator variables.

Forecasts computed using the extended population of models $M_1$ are reported as DMA-E. The results are very close to those of DMA-R. When there are differences in the assessment, these are not sizeable and completely disappear if a sufficient size for population $M_0$ is allowed. Intuitively, the population $M_1$ has the advantage of always including all regressors in its models and as a consequence it should react more quickly to model changes. On the other hand its forecast is slightly more noisy due to the presence of additional models. The two effects basically cancel out. DMA-R uses many fewer models than DMA-E and so may be preferred to DMA-E on the grounds of simplicity, ease of interpretation and computational efficiency.

Each DMA method beats the corresponding forecast computed with DMS, although by a small margin, corroborating the common finding that model averaging can beat even the best model in the pool.

We also implemented DMA using models with time varying parameters. 
In our case, this tended to overreact to the crisis, showing poorer performance. We interpret this result as hinting that changes in models during the crisis were not due to strong nonlinearities in the model but to the appearance of new regressors.

Following \citet{KoopKorobilis2012}, we tried additional benchmarks. 
These included a single time-varying parameter  model including all regressors, and  a single Bayesian ordinary least squares model with all regressors. 
However, these models either could not be estimated or performed very poorly 
as their estimated paramaters were very unstable.

\subsection{Further results}
\begin{figure}
   \centering
       \includegraphics[width=1\textwidth]{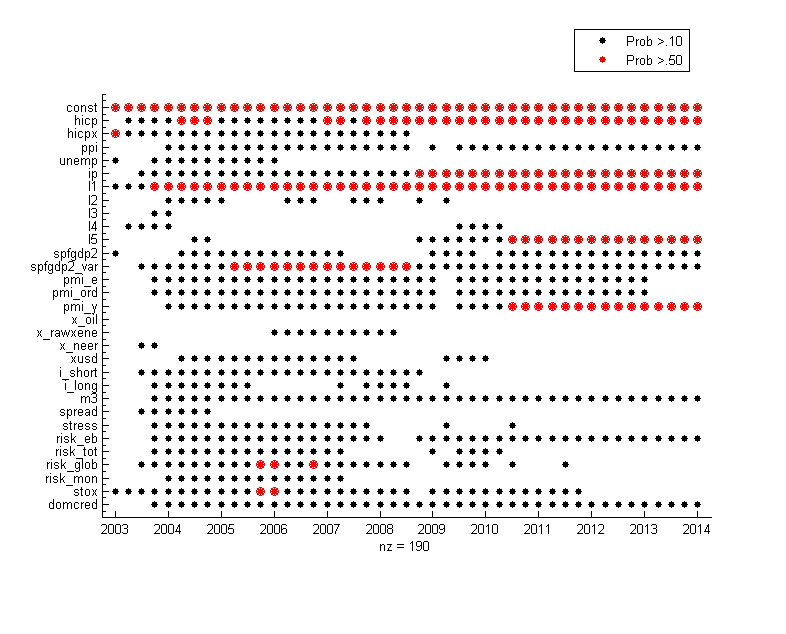}
   \caption{\label{fig-inclusionprobs} Inclusion probabilities of variables over time: (black) above 10\%, (red) above 50\%}
 \end{figure}

\begin{figure}
\begin{center}
       \includegraphics[width=1.08\textwidth]{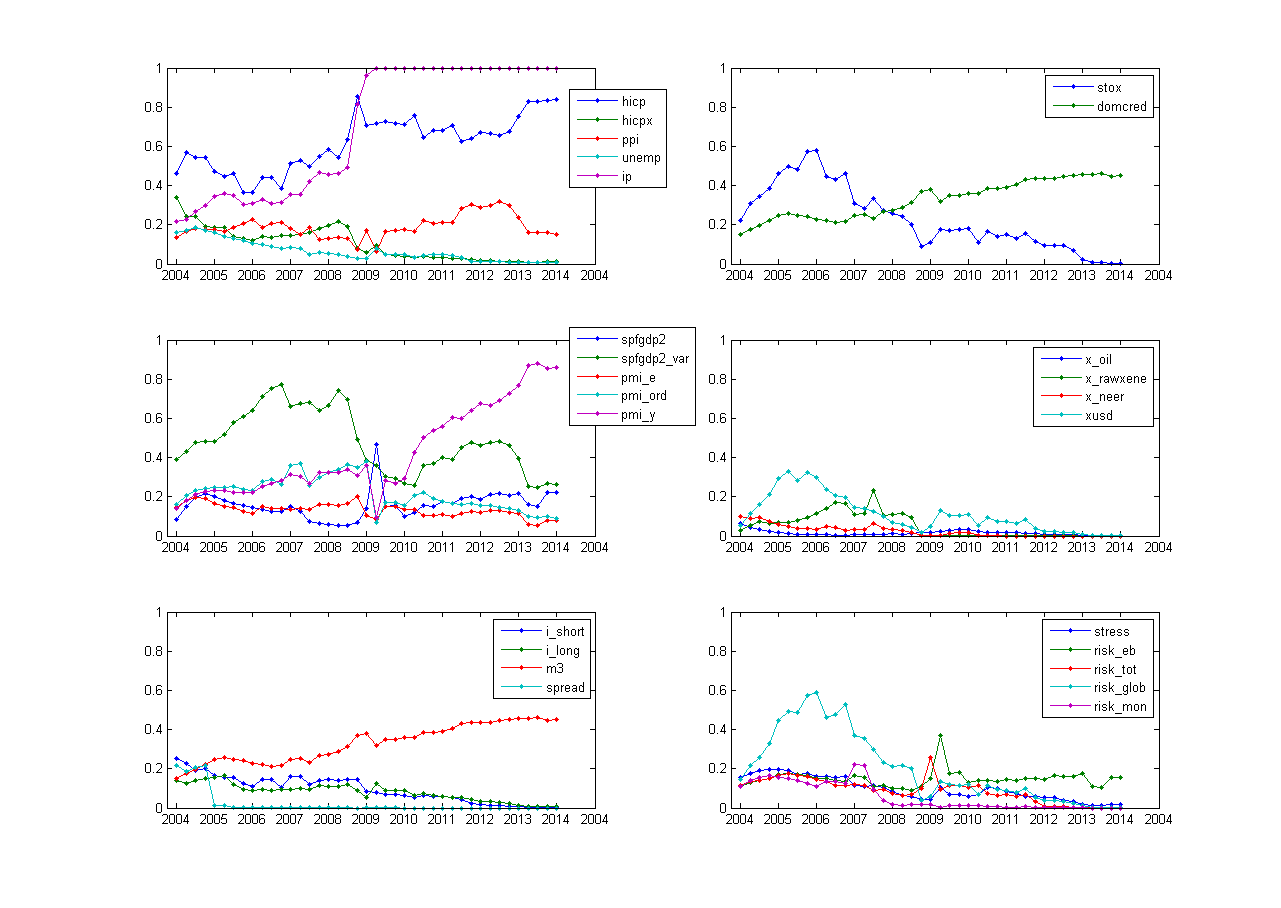}
   \caption{\label{fig-inclusionprobs.ts} Inclusion probabilities of single variables over time}
\end{center}
 \end{figure}

An important value added (beyond the good forecasting performance) of using model averaging concerns the inclusion probabilities of each regressor and their evolution. DMA identifies the importance of single variables and how this varies over time, 
which helps interpretation.
The inclusion probability of a variable at a given time point is calculated by summing up the weights of the models that use that variable as a regressor. 
Thus they vary between 0 and 1 and give a measure of the importance 
of that regressor.
Their evolution in time is summarized in Figures \ref{fig-inclusionprobs}
and \ref{fig-inclusionprobs.ts}.

The inclusion probabilities identify which were the most useful indicators of 
real activity and how this changed over time. In more detail:
\begin{itemize}
\item Lags of GDP were, as expected, important overall. The first lag captures the persistence in GDP, and it remained important even during the crisis, when GDP showed pronounced swings. The fourth and fifth lag capture essentially base effects. Our decision to include lag 5 as a potential regressor turned out to be justified.

\item Among the consumer price variables, HICP was an important regressor over the whole sample. This confirms the idea that prices and output are not determined in isolation. Without extending our interpretation to the existence of a European Phillips curve, we notice that these results confirms the results for the euro area recently obtained by \citet{Giannone&2014}. Furthermore, the DMA emphasized the role of producer prices as a forward-looking indicator for nowcasting GDP.

\item Among the early indicators of real activity, industrial production was the most important. This is a well known result in nowcasting, where industrial production is widely used as a timely and already comprehensive subset of GDP. The role of unemployment changed over time, becoming less important in the aftermath of the crisis. 

\item DMA selected almost all GDP surveys as important over the sample, with the exception of the period immediately following the 2008 crisis, which the surveys fail to capture adequately. This results confirm the literature on nowcasting inflation \citep{Ang&2007},
arguing that surveys have a relevant nowcasting power, thereby supporting the importance of expectations in determining macroeconomic outcomes. 

\item No single external variable alone had a determinant role. This is possibly due to the relative compactness of the euro area. Even variables traditionally important in determining prices, such as oil prices or the exchange rate, appear to have had a limited impact on real GDP. We find this result interesting but not surprising, given that these variables mostly affect prices, and affect GDP only indirectly. 

\item Among the variables closer to the operation of monetary policy, interest rates progressively lost their importance in the credit constrained post-crisis period, while the monetary variable M3 had an increasing role, possibly highlighting the importance of liquidity in the recent part of the sample.

\item Risk variables were useful predictors for GDP. Almost all risk indicators seemed to matter before the crisis, in particular the indicator of global risk.
The indicator of financial and banking risk remained important 
during the crisis.

\end{itemize}

\citet{KoopOnorante2014} carried out a similar analysis of inflation.
They used DMA on a similar dataset, but they explored the whole model space,
which limited the number of predictor variables they could use.
Comparing their results with ours, it is apparent that the determinants
of GDP growth were fairly similar to those of inflation.
The differences are largely intuitive: while cycle variables influence consumer prices, GDP is also well forecasted with producer prices, determined in advance; international prices do not have the same importance for the cycle as they have for inflation. On the other hand, as in \citet{KoopOnorante2014}, variables representing expectations are important predictors, with the exception of the crisis period.

\begin{figure}
\begin{center}
       \includegraphics[width=1.08\textwidth]{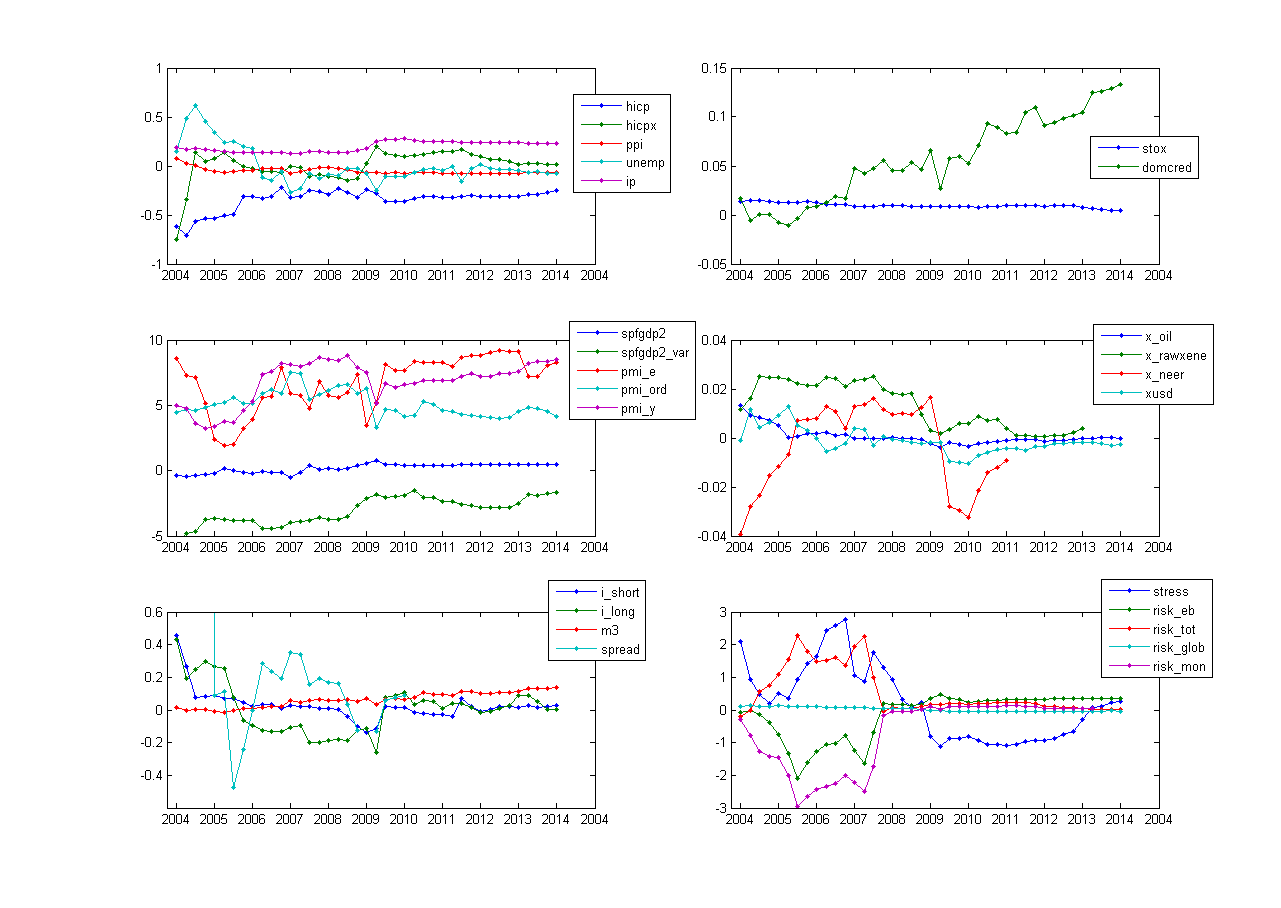}
   \caption{\label{fig-postmean} Posterior Means of Regression Coefficients
Over Time Under DMA.}
\end{center}
 \end{figure}

A natural complement to the results above is the average sizes of the coefficient of each variable. These are shown in Figure \ref{fig-postmean}. 
While inclusion probabilities provide important information about which variables should be included in the regressions at each point in time (as a significance test would do), they do not specify the size of their effect, and even a variable with a very high inclusion probability may have a small overall impact on GDP. The coefficients are averages over models at each point in time, and so vary over time. 

\section{Sensitivity Analysis}

\subsection{Initial conditions}

The DOW method requires the specification of an initial set of
models at the first time point. In our implementation we used an initial
set consisting of just one model, the constant model with no regressors,
or null model.
In this section we check the sensitivity of the forecast to the choice 
of the initial population of models.

\begin{figure}
   \centering
       \includegraphics[width=1\textwidth]{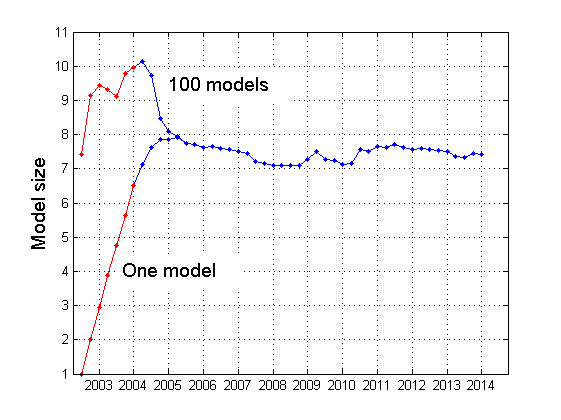}
   \caption{\label{fig-modelsize} Evolution of average model size starting from different initial model populations: one single model (the constant) or a random population of 100 models of average size 8}
 \end{figure}

Figure \ref{fig-modelsize} reports the average number of variables included in the models. The same average size of models is a necessary but not sufficient condition for convergence in populations of models, but it allows an easy graphic exploration of convergence.

It appears that DMA favours models with about 7-8 variables, and that the initial population $M_0$ is not representative of the final models selected. 
Figure \ref{fig-inclusionprobs} supports this finding by showing that the inclusion probabilities change rapidly at the beginning of the sample. Beyond the few periods of presample (in red) and a few initial points, two very different initial conditions: 1) an initial population of a model only, including only the constant, and 2) with a random population of 100 models, give very comparable results both in terms of model size (Figure \ref{fig-modelsize}) and forecasting power.

Table \ref{tbl-initialmodelsize8} compares the forecasting performance of 
different methods with various flavours of DMA and DMS, staring with a random
initial population of models with average size 8. The results are very
similar to those in Table \ref{tbl-performance}, in which DMA was 
initialized with the null model only. Once again, the DMA methods
outperformed the others considered. 

\begin{table}
   \centering
 \caption{\label{tbl-initialmodelsize8} Forecasting Results Starting From an 
Initial Population of Models of Average Size 8. The best performers under
each criterion are shown in bold.}
\begin{tabular}{@{}lllllll@{}}
\toprule
     & RW   & AR2  & DMA-R& DMA-E& DMS-R& DMS-E    \\ \midrule
RMSE & 0.0101 & 0.0088 & {\bf 0.0043} & {\bf 0.0043} & 0.0047 & 0.0047  \\
MAE  & 0.0067 & 0.0059 & {\bf 0.0032} & {\bf 0.0032} & 0.0035 & 0.0035  \\
MAX  & 0.0332 & 0.0376 & {\bf 0.0127} & {\bf 0.0127} & 0.0139 & 0.0139  \\ \bottomrule
\end{tabular}
\end{table}

\subsection{Maximum number of models}

\begin{figure}
   \centering
       \includegraphics[width=1\textwidth]{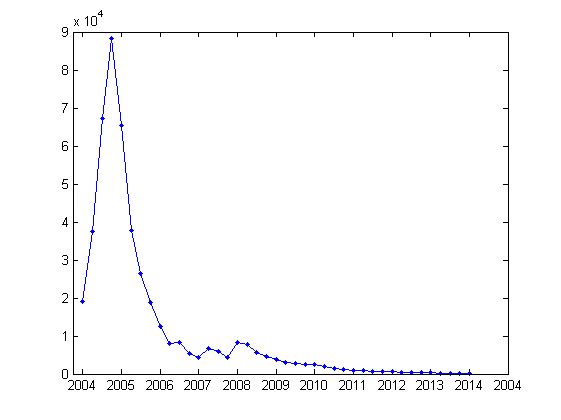}
   \caption{\label{fig-numberofmodels} Occam's window and number of models over time: $M_0$}
 \end{figure}

A pure application of the Occam's window principle and of the FEAR algorithm would require each model to satisfy condition (\ref{eq-OW}). This would soon lead to a relatively high number of models in the wider population $M_1$, as this population, generated from the Expand step of the algorithm, includes all possible neighbours of the preceding population $M_0$. The latter, however, is comparatively well contained by the following Reduce step, where condition (\ref{eq-OW}) is applied. 
Figure \ref{fig-numberofmodels}  shows the evolution of the size of population $M_0$ over time. 

The effort of the algorithm to find a stable population of models at the beginning is reflected in the high number of models retained. It is important to note that we start our evaluation sample after ten data points; as a result many models are poorly estimated at first and their performance varies considerably. After a few periods, a stable population has been found and it is progressively refined; as a result the size of $M_0$ decreases. 

Starting from a valid population, as in our hypothesis, the FEAR algorithm increases the population size only during turbulent times, for example at
the beginning of the Great Recession in 2008-2009. The algorithm automatically increases the population $M_0$ because the forecast is less accurate and no model is clearly dominating. This leads the FEAR algorithm to ``resuscitate'' additional models  in the attempt to improve the forecast. Figure \ref{fig-numberofmodels} shows that this attempt is usually successful. Quiet periods are instead characterized by smaller, stable model populations. 

Finally, as the sample size increases and models including the best regressors are selected the necessary population size becomes quite small (the last $M_0$ had size 186).
Overall, the population $M_0$, from which the baseline nowcast is generated, never exceeded 10,000 models, while the wider $M_1$ can be up to about ten times larger.

\begin{figure}
   \centering
       \includegraphics[width=1\textwidth]{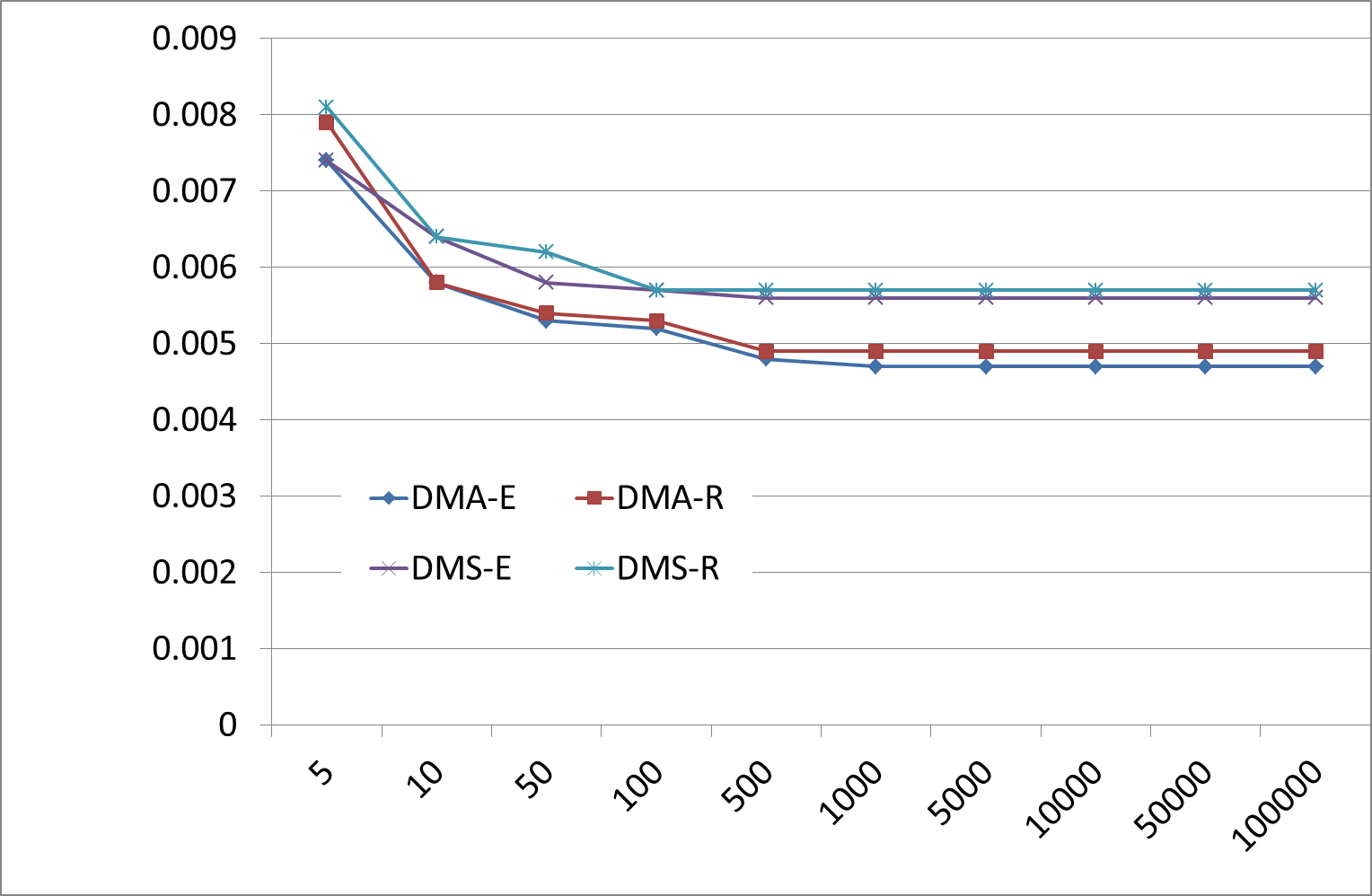}
   \caption{\label{fig-RMSE} Number of models and RMSE of nowcasting}
 \end{figure}

In the interest of speed, we introduced the possibility of specifying a maximum number of models $N$, and our last sensitivity analysis experiments with this number in order to assess whether it implies a deterioration of the forecast. 
Figure \ref{fig-RMSE} reports the nowcasting performance (as measured by the RMSE) in relation to maximum model size N. 

In our model space of 27 potential regressors, forecasting performance improved until about 10,000 models in the population $M_0$. 
Bigger model populations do not lead to any further improvement, 
as we have seen from the unconstrained estimation, 
and constraints set at 50,000 or above on the total population are not binding and thus exactly equivalent to Occam's window without a maximum number of models. We would of course still recommend keeping the maximum number of models as high as possible. 

Figure \ref{fig-RMSE} also confirms that in our case DMA performs slightly better than DMS for any population size. This is a robust result in the case of macroeconomic variables, but it cannot be generalized. 
\citet{KoopOnorante2014} and \citet{Morgan&2014},
for example, have shown using Google searches as predictors that DMS performed better in contexts where the data were noisy and forecasting benefitted from excluding many regressors. 

When looking at specific parameter values we observe that convergence may be slower for those parameters characterized by low inclusion probabilities. For some specific parameters and inclusion probabilities there are observable convergence issues up to 50,000 models. When this more specific information is important, we would suggest increasing the maximum number of models by one (or if possible two) orders of magnitude.

\section{Conclusions}
We have proposed a new method for carrying out Dynamic Model Averaging
when the model space is too large to allow exhaustive evaluation,
so the original DMA method of \citet{Raftery&2010} is not feasible.
This method, based on Occam's window and called Dynamic Occam's Window (DOW), 
is particularly efficient in situations in which numerous time series of limited length are available, as is typically the case in macroeconomics. Our procedure allows us to perform Dynamic Model Averaging without considering the whole model space but using a subset of models and dynamically optimizing the choice of models at each point in time.

We tested the model in an important empirical application, nowcasting GDP in the euro area. We showed that the forecasting performance was satisfactory compared to common benchmarks and that the results compare well with recent literature and with estimations performed on similar data sets. Several sensitivity analyses confirm the robustness of our approach to the choice of the user-specified
control parameters.

\paragraph{Acknowledgements:} Raftery's research was supported by
NIH grants R01 HD054511 and R01 HD070936 and by a Science Foundation 
Ireland E.T.S.~Walton visitor award, grant reference 11/W.1/I2079.
Raftery thanks the School of Mathematical Sciences at University College
Dublin for hospitality during the preparation of this article.

\bibliographystyle{chicago}
\bibliography{bigdma}

\end{document}